\documentclass[a4paper,fleqn]{cas-dc}

\usepackage[numbers, sort, compress]{natbib}
\usepackage{wasysym}
\usepackage{xcolor}
\usepackage{soul}

\def\tsc#1{\csdef{#1}{\textsc{\lowercase{#1}}\xspace}}
\tsc{WGM}
\tsc{QE}
\tsc{EP}
\tsc{PMS}
\tsc{BEC}
\tsc{DE}

\begin{document}
\let\WriteBookmarks\relax
\def\floatpagepagefraction{1}
\def\textpagefraction{.001}
\shorttitle{Percolation Transitions in Growing Networks Under Achlioptas Processes: Analytic Solutions}
\shortauthors{Soo Min Oh et~al.}

\title [mode = title]{Percolation Transitions in Growing Networks Under Achlioptas Processes: Analytic Solutions}
\tnotemark[1]

\tnotetext[1]{This research was supported by the National Research Foundation of Korea (NRF) through Grant Nos. NRF-2014R1A3A2069005 (B.K.) and NRF-2020R1A2C2010875 (S.W.S.) and by the TJ Park Science Fellowship from the POSCO TJ Park Foundation (S.W.S.).}


\author[1]{S. M. Oh}[alt=Soo Min Oh,
                        type=editor,
                        auid=000,bioid=1,
                        orcid=0000-0003-2186-1232
						]
\ead{dotoa@snu.ac.kr}

\credit{Conceptualization, Methodology, Software, Investigation, Formal analysis, and Writing - Original draft}

\address[1]{CCSS, CTP and Department of Physics and Astronomy, Seoul National University, Seoul 08826, Korea}

\author[2,3]{S.-W. Son}[alt=Seung-Woo Sow, orcid=0000-0003-2244-0376]

\cormark[1]
\ead{sonswoo@hanyang.ac.kr}

\credit{Conceptualization, Investigation, Writing - Review and editing, and Funding acquisition}

\address[2]{Department of Applied Physics, Hanyang University, Ansan 15588, Korea}
\address[3]{Asia Pacific Center for Theoretical Physics, Pohang 37673, Korea}

\author[1]{B. Kahng}[alt=Byungnam Kahng, orcid=0000-0002-9099-6395]
\cormark[1]
\ead{bkahng@snu.ac.kr}

\credit{Conceptualization, Investigation, Writing - Review and editing, Funding acquisition, and Supervision}

\cortext[cor1]{Corresponding authors}


\begin{abstract}
%
%
Networks are ubiquitous in diverse real-world systems. Many empirical networks grow as the number of nodes increases with time. Percolation transitions in growing random networks can be of infinite order. However, when the growth of large clusters is suppressed under some effects, e.g., the Achlioptas process, the transition type changes to the second order. However, analytical results for the critical behavior, such as the transition point, critical exponents, and scaling relations are rare. Here, we derived them explicitly as a function of a control parameter $m$ representing the suppression strength using the scaling ansatz. We then confirmed the results by solving the rate equation and performing numerical simulations. Our results clearly show that the transition point approaches unity and the order-parameter exponent $\beta$ approaches zero algebraically as $m \to \infty$, whereas they approach these values exponentially for a static network. Moreover, the upper critical dimension becomes $d_u=4$ for growing networks, whereas it is $d_u=2$ for static ones.
\end{abstract}

\maketitle

\section{Introduction}
Percolation describes the emergent behavior of connected clusters in complex networks~\cite{stauffer_introduction_2018, christensen_complexity_2005, albert_statistical_2002, dorogovtsev_evolution_2002, newman_structure_2003, boccaletti_complex_2006, dsouza_anomalous_2015, araujo_recent_2014}. It is well known that the Erd\H{o}s-R\'enyi (ER) random network model~\cite{erdos_evolution_1960} exhibits a second-order percolation transition at the critical probability $p_c$, where the order parameter $G$, which represents the fraction of nodes belonging to the giant cluster, behaves as $G \sim (p-p_c)^\beta$ with $\beta=1$. About a decade ago, various types of local suppression rules were proposed to alter the percolation transition type; these rules include product/sum rules~\cite{achlioptas_explosive_2009, dsouza_anomalous_2015}, the adjacent-edge rule~\cite{dsouza_local_2010, grassberger_explosive_2011}, and the da Costa rule~\cite{da_costa_explosive_2010,da_costa_solution_2014,da_costa_solution_2015}. These procedures, which are referred to as Achlioptas processes (APs)~\cite{achlioptas_explosive_2009}, locally suppress the growth of larger clusters but support that of smaller ones. Percolation transitions under APs were once believed to be discontinuous transitions at delayed transition points~\cite{achlioptas_explosive_2009}. However, it was verified that the percolation transitions in complex networks under local suppression rules are of the second order but have an extremely small critical exponent, i.e., $\beta \approx 0^+$, demonstrating the robustness of the second-order percolation transition~\cite{grassberger_explosive_2011, da_costa_explosive_2010, riordan_explosive_2011, lee_continuity_2011, cho_avoiding_2013}. All of the aforenoted information pertain to static networks, which have a fixed number of nodes. 

Numerous real-world examples exist regarding networks whose total number of nodes increases with time. Those examples include the World Wide Web and social networks~\cite{albert_statistical_2002, dorogovtsev_evolution_2002, newman_structure_2003, boccaletti_complex_2006, dsouza_anomalous_2015, araujo_recent_2014}. A simple growing random network (GRN) model, in which an infinite-order percolation transition occurs, has been introduced~\cite{callaway_are_2001, dorogovtsev_anomalous_2001, sole_model_2002, kim_infinite-order_2002}. In our recent report~\cite{oh_explosive_2016}, we proposed a minimal rule that locally suppresses the growth of large clusters in the GRN. First, a node is added to the system at each time step. Subsequently, $m$ candidate nodes among the present nodes are selected. Next, two nodes that belong to the two smallest clusters are connected by a link with probability $p$. This minimal rule is applicable to static network models, similar to the da Costa rule, which compares two sets of $m$ nodes and selects nodes belonging to the smallest cluster in each set. The difference between the two rules is that the minimal rule requires only a single comparison, whereas the da Costa rule requires double comparisons. By numerically solving the rate equation of the cluster size distribution~\cite{ziff_kinetics_1983, leyvraz_scaling_2003, cho_cluster_2010} under the minimal rule, which resembles the Smoluchowski equation~\cite{smoluchowski_uber_1916}, we discovered that the infinite-order percolation transition became a second-order percolation transition~\cite{oh_explosive_2016, yi_percolation_2013}. Let us denote this model as $m$-GRN. Similarly, $m$-ER denotes the ER model under the minimal rule, and $d$-ER and $d$-GRN represent static and growing random networks, respectively, under the da Costa rule~\cite{da_costa_explosive_2010}.
 
In both the $m$- and $d$-GRN models, as $m$ is increased, the suppression effect becomes stronger, and the percolation transitions occur explosively with an extremely small value of the critical exponent $\beta$. Phase transitions appear abruptly, but they are still continuous. This is true even for $m$- and $d$-ER static models with limited information regarding the local suppression~\cite{grassberger_explosive_2011,da_costa_explosive_2010,oh_explosive_2016}. By contrast, when a global suppression rule using global information is applied, as in restricted ER~\cite{cho_hybrid_2016} and restricted GRN models~\cite{oh_suppression_2018, oh_discontinuous_2019}, the percolation transition becomes a hybrid transition~\cite{cho_hybrid_2016,lee_recent_2018} and a first-order discontinuous transition~\cite{oh_suppression_2018, oh_discontinuous_2019}, respectively. These types of discontinuous percolation transitions appear in interdependent networks~\cite{buldyrev_catastrophic_2010, baxter_avalanche_2012, son_percolation_2012, zhou_simultaneous_2014, havlin_percolation_2015, cellai_message_2016, lee_universal_2017}, and in sparse networks~\cite{bianconi_rare_2018} when a large deviation of the giant component size is considered. Furthermore, it was confirmed that a discontinuous percolation transition occurred at a nontrivial critical point in one-dimensional~\cite{araujo_recent_2014} and two-dimensional lattices~\cite{choi_critical_2017}, and on Farey graphs~\cite{boettcher_ordinary_2012} and hyperbolic manifolds~\cite{kryven_renormalization_2019} when certain long-range interactions were considered.

Here, we analytically investigated the scaling relations of the critical exponents in the GRN under local suppression rules. Although several studies have been performed to investigate the critical behavior of the percolation model under APs, analytic solutions were applicable to only a few static cases~\cite{da_costa_explosive_2010,da_costa_solution_2014,da_costa_solution_2015}. Enabled by the analytical tractability of the da Costa rule, we first derived the scaling relations for the $d$-GRN model following the approach used in ~\cite{da_costa_explosive_2010,da_costa_solution_2014,da_costa_solution_2015}. Next, we analyzed the results for the $d$-GRN model and compared them, together with the $d$-ER result, with those of the $m$-GRN and $m$-ER models. Subsequently, we used $P_m(s,p,t)$, the probability that a selected node belongs to the cluster of size $s$; $Q_m(s,p,t)$, the probability that the smallest cluster among the $m$ clusters to which $m$ randomly selected nodes belong at time $t$ is of size $s$ for a specified link connection probability $p$; and a control parameter $m$ defined in the da Costa rule. Assuming that $P_m(s,p,t)$ and $Q_m(s,p,t)$ have scaling functions in the steady state limit, we obtained similar scaling relations for the critical exponents in percolation theory~\cite{stauffer_scaling_1979}. We then confirmed these results by solving the rate equations numerically~\cite{oh_explosive_2016}.

This paper is organized as follows. We introduce the model and derive the rate equation of $P_m(s,p,t)$ in Section~\ref{sec:model and rate equation}. The scaling relations of the critical exponents are derived using the scaling functions $P_m(s,p,t)$ and $Q_m(s,p,t)$ in Section~\ref{sec:Scaling functions and scaling relations}. The transition points $p_{c}$ are derived for general values of $m$ in Section~\ref{sec:equations of scaling functions}. We solve the rate equation of $P_m(s,p,t)$ numerically and confirm our main results in Section~\ref{sec:Numerical solution of rate equation}. The hyperscaling relations are discussed near the end of this section. The results of this study are summarized, and their implications are discussed in Section~\ref{sec:summary and discussion}.

\section{Model and rate equation}
\label{sec:model and rate equation}
\begin{figure}[]
\center
\includegraphics[width=0.8\linewidth]{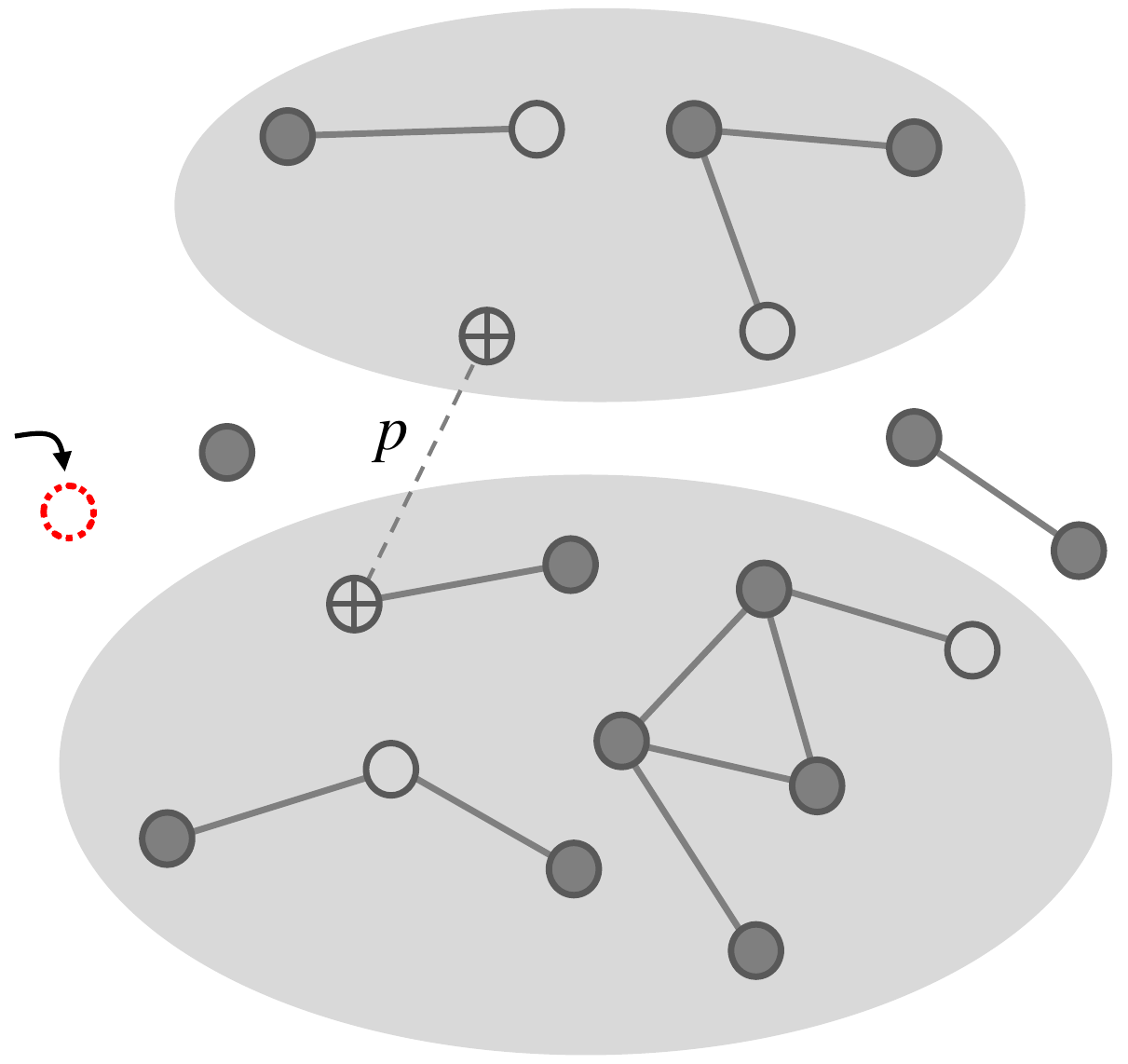}
\caption{(Color online) 
Schematic illustration of our model for $m=3$. Nodes are represented by solid circles. After a new node (red dotted open circle) is added to the system, two sets of $m$ nodes (solid open circles) are randomly selected from distinct clusters. For each set represented by a shaded ellipse, the node belonging to the smallest cluster (represented by $\oplus$) is chosen. The two nodes $\oplus$ are connected with probability $p$ by a link (dashed line).
}
\label{fig:Schematic_figure}
\end{figure}

We consider a growing network model under a rule that suppresses the growth of large clusters locally with limited information. It consists initially of an isolated node, and a new node is added to the system at each time step; consequently, the total number of nodes at time $t$ is $N(t) = t + 1$. Then, two sets of $m$ candidate nodes are selected randomly. The node that belongs to the smallest cluster in each set is selected, and these two nodes are connected with the wiring probability $p$, as depicted schematically in Fig.~\ref{fig:Schematic_figure}. When $m = 1$, this growing network model reduces to the GRN model proposed by Callaway et al.~\cite{callaway_are_2001}. This type of suppression rule in static network models was first considered by da Costa et al.~\cite{da_costa_explosive_2010}. When $m = 1$, this da Costa model also becomes the ER random network model~\cite{erdos_evolution_1960}. The similar, but simpler, minimal rule is considered for the growing and static network models~\cite{oh_explosive_2016} by applying the local suppression rule, where two nodes belonging to the two smallest clusters among $m$ randomly selected nodes are connected with probability $p$. In the unified framework, we derive the analytic solutions of all these models for growing and static networks.

Adopting the notation of the da Costa model~\cite{da_costa_explosive_2010, da_costa_solution_2014}, we define $P_m(s,p,t)$ as the probability that a selected node belongs to the cluster of size $s$ at time $t$ for a given control parameter $m$ representing the strength of suppression, where $p$ denotes the probability that a link is added between the two selected nodes. Then the rate equation of $P_m(s,p,t)$ is written as 
\begin{align}
\label{eq:rate_eq_of_P_s_p_t}
&\frac{d}{dt} \Bigl( N(t) P_m(s,p,t) \Bigr) \\
&= sp \Bigl[\sum_{u+v=s}Q_m(u,p,t)Q_m(v,p,t) - 2Q_m(s,p,t) \Bigr] \nonumber \\
&+ \delta_{1s} \nonumber ,
\end{align}
where
\begin{align}
\label{eq:Q_p_t}
&Q_m(s,p,t) \\
&= \sum_{k=1}^{m} \binom{m}{k} P_m(s,p,t)^{k} \Bigl[ 1-\sum_{u=1}^{s}P_m(u,p,t) \Bigr]^{m-k} \nonumber .
\end{align}

$Q_m(s,p,t)$ is the probability that the smallest cluster among the $m$ clusters to which the $m$ randomly chosen nodes belong at time $t$ is of size $s$ for a given $p$. The last term, $\delta_{1s}$, in Eq.~\eqref{eq:rate_eq_of_P_s_p_t} indicates that a new node of size one is added to the system at each time step. For static networks, the last term disappears, and the total number of nodes $N(t)$ is fixed at constant $N$. Moreover, the linking probability $p$ is unity because a link is always added at each time step in the static network model. The above rate equation of $P_m(s,p,t)$ is equivalent to that in Refs.~\cite{da_costa_explosive_2010, da_costa_solution_2014} with the time normalized by the system size $N$. 

\section{Scaling relations of critical exponents}
\label{sec:Scaling functions and scaling relations}
Here, we try to determine the scaling relations using the scaling forms of $P_m(s,p,t)$ and $Q_m(s,p,t)$ in growing networks. In the steady state limit, i.e., $N \gg 1$ and $t \gg 1$, assuming that $P_m(s,p,t)$ and $Q_m(s,p,t)$ are independent of time $t$, they thus can be written as $P_m(s,p)$ and $Q_m(s,p)$. Then Eq.~\eqref{eq:rate_eq_of_P_s_p_t} becomes
\begin{align}
\label{eq:rate_eq_of_P_s_p}
&P_m(s,p) \\
&= sp \Bigl( \sum_{u+v=s}Q_m(u,p)Q_m(v,p) - 2Q_m(s,p) \Bigr) + \delta_{1s} \nonumber.
\end{align}

As $p$ is increased, cluster formation becomes more likely. Numerical simulations~\cite{oh_explosive_2016} show that, above the percolation threshold $p_c$, a percolating cluster of size $G$ emerges as $G \sim (p-p_c)^\beta$ for $m \ge 2$. The two distributions, $P_m(s,p)$ and $Q_m(s,p)$, satisfy the sum rules $\sum_s P_m(s,p) = 1 - G$ and $\sum_s Q_m(s,p) = 1 - G^m $, where an infinite cluster is excluded from the sums. The $n$-th moments of the cluster sizes for each distribution are expressed as $\langle s^n \rangle_P = \sum_{s}s^n P_m(s,p)$ and $\langle s^n \rangle_Q = \sum_{s}s^n Q_m(s,p)$. Eq.~\eqref{eq:rate_eq_of_P_s_p} for finite components leads to the following equations:
\begin{align}
\label{eq:<s>_q}
G &= 2p G^{m} \langle s \rangle_Q,  \\
\label{eq:<s>_p}
\langle s\rangle_P &= 2p\langle s\rangle_Q^{2}-2p G^{m}\langle s^2 \rangle_Q + 1.
\end{align}

Next, $P_m(s,p)$ is assumed to follow scaling behavior near $p_{c}$ as  
\begin{align}
\label{eq:P_s_p_t}
P_m(s,p)=s^{1-\tau}f(s/s_c),
\end{align}
where $s_c$ is a characteristic cluster size and behaves as $s_c \sim |p-p_c|^{-1/\sigma}$. In addition, $f(x)$ is a scaling function that by definition is constant for $x \ll 1$ and decays exponentially for $x \gg 1$. From this, we obtain that $\beta=(\tau-2)/\sigma$.

Replacing the summation in Eq.~\eqref{eq:Q_p_t} with an integral, we find
\begin{align}
\label{eq:Q_s_p_1}
Q_m(s,p) \cong m \Bigl( \int_{s}^{\infty}du P_m(u,p) \Bigr)^{m-1} P_m(s,p) 
\end{align}
for large $s$ in the steady state limit. Then the scaling form of $Q_m(s,p)$ is obtained as follows:
\begin{align}
\label{eq:Q_s_p_2}
Q_m(s,p)=s^{(2m-1)-m\tau} g(s/s_c),
\end{align}
where $g(x)$ is a scaling function of $Q_m(s,p)$, corresponding to $f(x)$ for $P_m(s,p)$. 

Because the first moments of the cluster sizes diverge at the critical point as $\langle s \rangle_P \sim (p-p_c)^{-\gamma_P}$ and $\langle s \rangle_Q \sim (p-p_c)^{-\gamma_Q}$, Eqs.~\eqref{eq:P_s_p_t} and \eqref{eq:Q_s_p_2} produce the following two scaling relations:
\begin{align}
\label{eq:gamma_P}
&\gamma_P = (3-\tau)/\sigma , \\
\label{eq:gamma_Q}
&\gamma_Q =  (2m+1-m\tau)/\sigma .
\end{align}
Moreover, plugging $\langle s \rangle_P$ and $\langle s \rangle_Q$ into Eqs.~\eqref{eq:<s>_q} and \eqref{eq:<s>_p}, we obtain that 
\begin{align}
\label{eq:gamma_rel}
\gamma_P = 2\gamma_Q = 2(m-1)\beta .
\end{align}
By using Eqs.~\eqref{eq:gamma_P}--\eqref{eq:gamma_rel}, the explicit forms of the critical exponents $\gamma_{P}$, $\gamma_{Q}$, $1/\sigma$, and $\tau$ are obtained in terms of $\beta$ and $m$ as follows: 
\begin{align}
\label{eq:scaling_rel_of_gamma_P}
&\gamma_P=2(m-1)\beta ,\\
\label{eq:scaling_rel_of_gamma_Q}
&\gamma_Q=(m-1)\beta ,\\
\label{eq:scaling_rel_of_sigma}
&\frac{1}{\sigma}=(2m-1)\beta ,\\
\label{eq:scaling_rel_of_tau}
&\tau=2+\frac{1}{2m-1} .
\end{align}

We remark that these formulas differ from the corresponding formulas for static network~\cite{da_costa_solution_2014}. The two exponent formulas for the static and growing cases are compared in Table~\ref{table_3}. 
We also note that the four formulas above are consistent with those obtained in the previous study~\cite{oh_explosive_2016} of the minimal rule, but $m$ is replaced by $2m$, because the minimal rule chose $m$ nodes randomly, and not $2m$ nodes as in this model. Finally, we remark that the exponent $\tau$ is independent of $\beta$ for the growing model but depends on $\beta$ for the static model.

In the supercritical regime, $p > p_c$, where the giant cluster emerges, Eq.~\eqref{eq:Q_s_p_1} can be simply approximated as $Q_m(s,p) \cong mG^{m-1}P_m(s,p)$. The generating functions of $P_m(s,p)$ and $Q_m(s,p)$ are introduced as 
$$\mathcal{P}_m(z,p) \equiv \sum_{s=1}^{\infty}P_m(s,p)z^s$$ and 
$$\mathcal{Q}_m(z,p) = \sum_{s=1}^{\infty}Q_m(s,p)z^s,$$ respectively. The relation between the two generating functions can be written as
\begin{align}
\label{eq:above_transition_point_org}
1-G^{m}-\mathcal{Q}_m(z,p) &= \sum_{s} Q_m(s,p)\Bigl[1-z^{s} \Bigr] \\
& \cong \sum_{s} mG^{m-1}P_m(s,p) \bigl[ 1-z^{s} \bigr] \nonumber \\
& = mG^{m-1} \bigl[ 1-G-\mathcal{P}_m(z,p) \bigr] \nonumber .
\end{align}
Therefore,
\begin{align}
\label{eq:above_transition_point}
1-\mathcal{Q}_m(z,p) = mG^{m-1} \Bigl[1-\mathcal{P}_m(z,p)-\frac{m-1}{m}G \Bigr] ,
\end{align}
where the sum rules $\sum_{s}Q_m(s,p) = 1- G^m$ and $\sum_{s}P_m(s,p) = 1- G$ are applied. Then, Eq.~\eqref{eq:rate_eq_of_P_s_p} becomes
\begin{align}
\label{eq:above_transition_point_2}
\mathcal{P}_m(z,p) &= 2m^2G^{2(m-1)}p\\ 
& \times \Bigl[\mathcal{P}_m(z,p) - 1 + \frac{m-1}{m}G \Bigr]\frac{\partial \mathcal{P}_m(z,p)}{\partial \ln z} + z . \nonumber 
\end{align}
When $z = 1$, the equation is 
\begin{align}
\label{eq:above_transition_point_3}
\mathcal{P}_m(1,p)-1 &= 2m^2G^{2(m-1)}p \\
& \times \Bigl[\mathcal{P}_m(1,p) - 1 + \frac{m-1}{m}G \Bigr]{\langle s \rangle}_{P}. \nonumber
\end{align}
Using the relations $G(p) \sim {(p-p_c)}^{\beta}$ and ${\langle s \rangle}_{P} \sim {(p-p_c)}^{-\gamma_P}$, one obtains $\gamma_P = 2(m-1)\beta$ again, which is consistent with Eq.~\eqref{eq:scaling_rel_of_gamma_P}.

\section{Analytic solution of the transition point}
\label{sec:equations of scaling functions}
To determine the transition point $p_{c}$, we derive the scaling functions of $f(x)$ and $g(x)$ with respect to $x$. First, by substituting $Q_m(u,p) = Q_m(s,p) + [Q_m(u,p) - Q_m(s,p)]$ into Eq.~\eqref{eq:rate_eq_of_P_s_p}, one obtains
\begin{align}
\label{eq:rate_eq_of_P_s_p_for_some_tricks}
&P_m(s,p) \\
&= p \Bigl[ -s(s-1)Q^2(s) + 2sQ_m(s) {\bigl(1-\sum_{u=s}^{\infty}Q_m(u)\bigr)} \nonumber \\
& + s\sum_{u=1}^{s-1}{\bigl(Q_m(u)-Q_m(s)\bigr)\bigl(Q_m(s-u)-Q_m(s)\bigr)} \nonumber \\
& - 2sQ_m(s) \Bigr] + \delta_{1s} \nonumber.
\end{align}

In the integral form for large $s$, this equation becomes 
\begin{align}
\label{eq:rate_eq_of_P_s_p_for_some_tricks_2}
&P_m(s,p) \\
&\cong p \Bigl[ -{s^2}Q^2(s) - 2sQ_m(s) \int_{s}^{\infty}Q_m(u) du  \nonumber \\
& + s\int_{0}^{s}{\bigl(Q_m(u)-Q_m(s)\bigr)\bigl(Q_m(s-u)-Q_m(s)\bigr) du} \Bigr] \nonumber. 
\end{align}

The scaling form of $P_m(s,p)$ for large $s$ in the critical region is $P_m(s,p)=s^{1-\tau}f(s{\delta}^{1/\sigma}) = {\delta}^{(\tau - 1)/{\sigma}} \tilde{f}(s{\delta}^{1/\sigma})$, where $\delta = |p - p_c| \ll 1$. In addition, the scaling form of $Q_m(s,p)$ is $Q_m(s,p) = s^{(2m-1)-m\tau} g(s{\delta}^{1/\sigma}) = {\delta}^{[m\tau-(2m-1)]/\sigma} \tilde{g}{(s{\delta}^{1/\sigma}})$. We obtain the following equation for the scaling functions.
\begin{align}
\label{eq:rate_eq_of_P_s_p_for_some_tricks_3}
\tilde{f}(x) & = p_c \Bigl[ -x^2{\tilde{g}}^2(x) - 2x\tilde{g}(x)\int_{x}^{\infty}{dy}{\tilde{g}(y)} \nonumber \\
& + x\int_{0}^{x}{dy}{[\tilde{g}(y) - \tilde{g}(x)][\tilde{g}(x-y) - \tilde{g}(x)] \Bigr]} ,
\end{align}
where $x\equiv s{\delta}^{1/\sigma}$, and $(2m-1)(\tau - 2) = 1$. This relation is also consistent with Eq.~\eqref{eq:scaling_rel_of_tau}. Using Eqs.~\eqref{eq:P_s_p_t} and \eqref{eq:Q_s_p_2}, we can obtain the following equation:
\begin{align}
\label{eq:rel_between_f_and_g}
\tilde{g}(x) = m { \Bigl[ \int_{x}^{\infty}{dy \tilde{f}(y) \Bigr]} }^{m-1} \tilde{f}(x) ,
\end{align}
where $g(x) = x^{m\tau-(2m-1)}\tilde{g}(x)$, and $f(x) = x^{\tau-1} \tilde{f}(x)$. These relations are all valid for the normal phase, $p < p_{c}$. For the percolating phase, $p > p_{c}$, Eqs.~\eqref{eq:rate_eq_of_P_s_p_for_some_tricks_3} and \eqref{eq:rel_between_f_and_g} are valid after the signs of each term that contains $\tilde{f}(x)$ are reversed.

Now, we assume that $f(x)$ and $g(x)$ are expandable for small $x$ around $0$ as follows:
\begin{align}
\label{eq:Expansion_of_f}
f(x) & = f(0) + a_{1}x^{\sigma} + a_{2}x^{2\sigma} + \cdots , \\
\label{eq:Expansion_of_g}
g(x) & = g(0) + b_{1}x^{\sigma} + b_{2}x^{2\sigma} + \cdots .
\end{align}

When Eqs.~\eqref{eq:Expansion_of_f} and \eqref{eq:Expansion_of_g} are substituted into Eqs.~\eqref{eq:rate_eq_of_P_s_p_for_some_tricks_3} and \eqref{eq:rel_between_f_and_g}, the relation between $f(0)$ and $g(0)$ becomes  
\begin{align}
\label{eq:Taylor_Series_expansion}
f(0) + O(x^{\sigma}) &= p_{c} \bigl( \frac{g^{2}(0) \Gamma[-m(\tau-2)]^2}{\Gamma[-2m(\tau-2)]} + O'(x^{\sigma}) \bigr) , \\
\label{eq:relation_between_f(0)_and_g(0)}
g(0) &= \frac{m}{(\tau-2)^{m-1}} f^{m}(0),
\end{align}
where $O(x^{\sigma})$ and $O'(x^{\sigma})$ represent the higher-order terms of $x^{\sigma}$. Unlike the equation for static networks, Eq.~\eqref{eq:rate_eq_of_P_s_p} contains the factor $p$ explicitly. Thus, the transition point $p_{c}$ can be determined by comparing the zeroth-order term of Eq.~\eqref{eq:Taylor_Series_expansion} together with  Eq.~\eqref{eq:relation_between_f(0)_and_g(0)} as follows:
\begin{align}
\label{eq:transition point}
p_{c} = \frac{f(0)^{1-2m}\Gamma [-2m(\tau-2)]}{m^2 (\tau-2)^{2-2m} \Gamma [-m(\tau-2)]^2},
\end{align}
where $\Gamma(z)$ is a gamma function defined as $\Gamma(z) \equiv (z-1)!$. When Eq.~\eqref{eq:scaling_rel_of_tau} is substituted into  Eq.~\eqref{eq:transition point}, the dependence of the critical exponent $\tau$ disappears and $p_c$ is obtained as follows:
\begin{align}
\label{eq:transition point_with_no_tau_dependence}
p_{c} &= \frac{f(0)^{1-2m}\Gamma [-2m/(2m-1)]}{m^2 (2m-1)^{2m-2} \Gamma [-m/(2m-1)]^2} \nonumber \\
&= \frac{f(0)^{1-2m}}{2m (2m-1)^{2m-2}} {B^{-1}\Bigl[ \frac{m-1}{2m-1},\frac{m-1}{2m-1} \Bigr]},
\end{align}
where $B^{-1}(x,y)$ is the inverse of the beta function and follows the relationship $B^{-1}(x,y)=\Gamma(x+y)/[\Gamma(x)\Gamma(y)]$. The transition point thus depends only on $f(0)$ and $m$. As shown in Ref.~\cite{da_costa_critical_2014}, the value of $f(0)$ can be estimated. Assuming that $P_m(s,p_{c})$ follows a power-law function such as $P_m(s,p_{c}) \sim f(0)s^{1-\tau}$ for all cluster sizes $s$ larger than another characteristic size $s_{0}$, we can write the normalization by $P_m(s,p)$ as follows.
\begin{align}
\label{eq:Find_f0}
\sum_{s=1}^{\infty}{P_m(s,p_{c})} = \sum_{s<s_{0}}{P_m(s,p_{c})} + f(0)\sum_{s \ge s_{0}}^{\infty}{s^{1-\tau}} = 1.
\end{align}

\begin{figure}[]
\center
\includegraphics[width=1.0\linewidth]{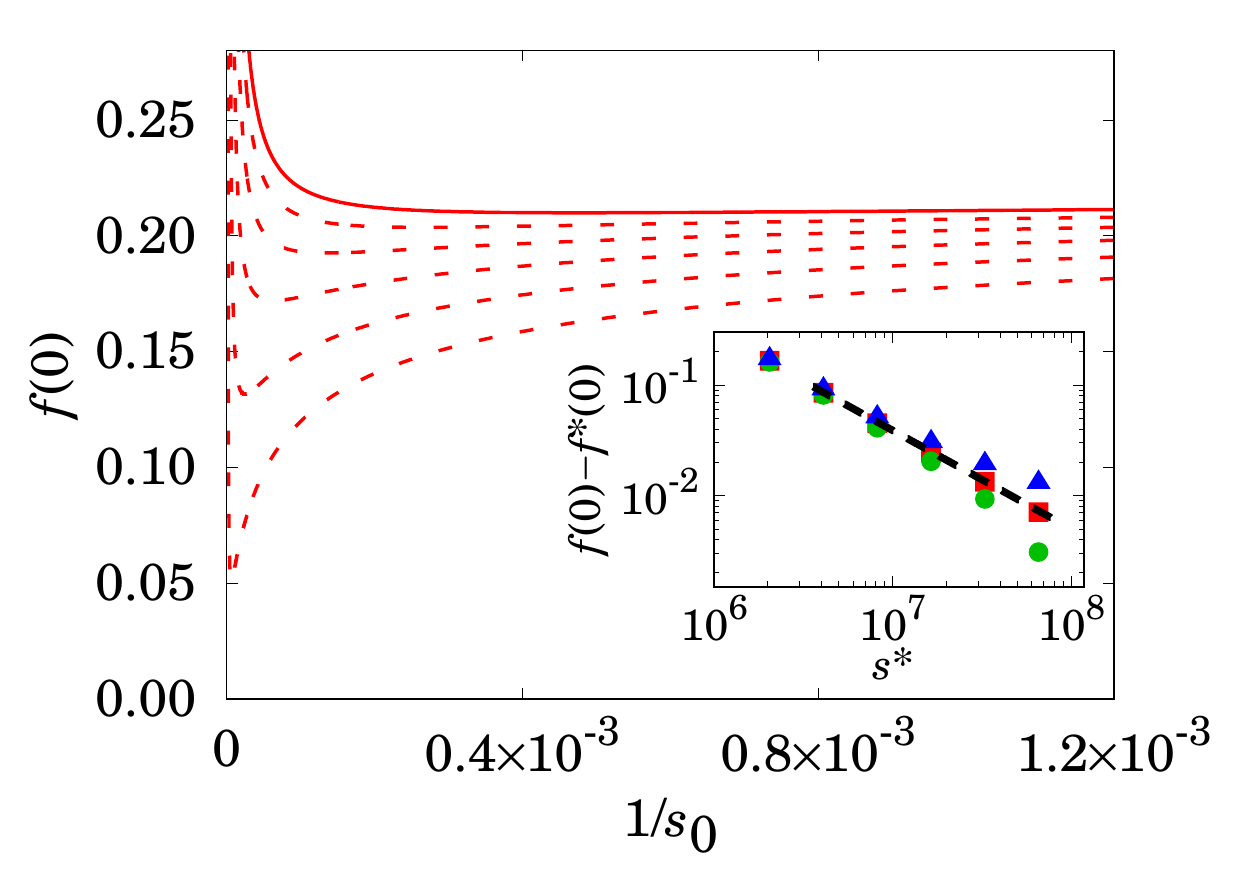}
\caption{(Color online) 
Plot of $f(0)$ versus $1/s_{0}$ for $m=2$ with $s_{0}$ for truncated cluster sizes $s^{*}/10^{3}= 2^{11}$, $2^{12}$, $2^{13}$, $2^{14}$, and $2^{15}$ (red dashed curves from bottom to top), and $2^{16}$ (red solid curve). Inset: The corresponding plot of $f(0)-f^{*}(0)$ versus $s^{*}$ at $p=0.51515$ around the transition point $p_{c}$, where $f^{*}(0)$ is the minimum value of $f(0)$ for a given $s^{*}$. Data points are for trial values of $f^{*}(0)=0.213$ ($\textcolor{LimeGreen} \CIRCLE$), $0.217$ ($\textcolor{red} \blacksquare$), and $0.223$ ($\textcolor{blue} \blacktriangle$). The black dashed line is a guide line and $f(0)$ is estimated as $0.217(1)$ for $m=2$.
}
\label{fig:f0_vs_inverse_s0}
\end{figure}

To solve Eq.~\eqref{eq:Find_f0} for $m=2$, we plot $f(0)$ versus $1/s_{0}$ in Fig.~\ref{fig:f0_vs_inverse_s0} and estimate $f(0)$ to be $\approx 0.217(1)$; ${p_{c}}^{*}=0.515(1)$. For various $m$ values between $2$ and $10$, $p_{c}^*$ is obtained using Eq.~\eqref{eq:transition point}; the results are listed with the corresponding $f(0)$ values in Table~\ref{table_1}. They are consistent with those obtained from the rate equations within the error bars. Moreover, we find that $f(0)$ decays asymptotically as $1/(2m+0.15)$.

Now, to investigate the asymptotic behavior of the percolation transition point $p_{c}$, we consider the Taylor expansion of Eq.~\eqref{eq:transition point_with_no_tau_dependence} around $1/m = 0$. For $m \gg 1$, we can use the approximations $B^{-1}\Bigl[\frac{m-1}{2m-1},\frac{m-1}{2m-1} \Bigr] \approx 1/\pi-(\gamma+\psi(1/2))/2\pi m$
and $(2m+0.15)^{2m-1}/[2m(2m-1)^{2m-2}] \approx ( 3.14 - 2.62/m )$, where $\gamma$ is the Euler-Mascheroni constant, and $\psi(z)$ is the zeroth-order polygamma function following the relation $\psi(z)=\Gamma'(z)/\Gamma(z)$. Substituting these approximations into Eq.~\eqref{eq:transition point_with_no_tau_dependence}, we derive the asymptotic behavior of $1-p_{c} \sim 1/m$, which decreases algebraically as $m$ is increased.

\begin{table*}[]
\centering
\caption{Numerical estimates of the percolation threshold $p_c$, the exponent of the order parameter $\beta$, the exponent of the cluster size distribution 
$\tau$, the exponent of the characteristic cluster size $\sigma$,
and the exponents of the susceptibility $\gamma_P$, $\gamma_Q$ 
of growing network models for $m=2$ to $10$. The transition point $p_{c}$ obtained from the rate equations is compared with ${p_{c}}^{*}$, which was analytically solved using the scaling functions. $f(0)$ is the coefficient of the leading term of $P_m(s,p) \approx f(0)s^{-\tau}$ for large $s$. We confirm that $p_{c}$ and ${p_{c}}^{*}$ are consistent with each other within errors.}
\label{table_1}
\begin{tabular}{ccccccccc}
\hline\hline
~~~~$m$~~~~  &~~~~ $f(0)$~~~~&~~~~ ${p_{c}}^{*}$~~~~&~~~~ $p_{c}$~~~~ &~~~~ $\beta$~~~~ &~~~~ $\tau$~~~~ &~~~~ $\sigma$~~~~ &~~~~  $\gamma_P$~~~~ &~~~~ $\gamma_Q$~~~~ \\ \hline
2 & 0.217(1) & 0.515(1) & 0.515(1)  & 0.457(1)   & 2.333(1)  & 0.730(2)  & 0.914(2)  & 0.458(1) \\
3 & 0.157(1) & 0.666(1) & 0.667(1)  & 0.242(1)   & 2.200(1)  & 0.827(2)  & 0.969(3)  & 0.484(1) \\
4 & 0.120(1) & 0.745(1) & 0.747(1)  & 0.164(1)   & 2.143(1)  & 0.872(2)  & 0.984(3)  & 0.492(1) \\ 
5 & 0.097(1) & 0.795(1) & 0.796(1)  & 0.124(1)   & 2.111(1)  & 0.898(2)  & 0.990(2)  & 0.495(1) \\ 
6 & 0.082(1) & 0.830(1) & 0.830(1)  & 0.099(1)   & 2.091(1)  & 0.916(2)  & 0.993(3)  & 0.497(1) \\ 
7 & 0.070(1) & 0.854(1) & 0.853(1)  & 0.083(1)   & 2.077(1)  & 0.928(2)  & 0.995(3)  & 0.498(1) \\
8 & 0.062(1) & 0.868(2) & 0.871(1)  & 0.071(1)   & 2.067(1)  & 0.937(2)  & 0.996(2)  & 0.498(1) \\ 
9 & 0.055(1) & 0.881(3) & 0.885(1)  & 0.062(1)   & 2.059(1)  & 0.944(2)  & 0.997(2)  & 0.499(1) \\ 
10 & 0.050(1) & 0.900(2) & 0.897(1)  & 0.055(1)   & 2.053(1)  & 0.950(2)  & 0.998(2)  & 0.499(1) \\ \hline 
\end{tabular}
\end{table*}

\section{Numerical solutions of the rate equation}
\label{sec:Numerical solution of rate equation}

\begin{figure}[]
\center
\includegraphics[width=1.0\linewidth]{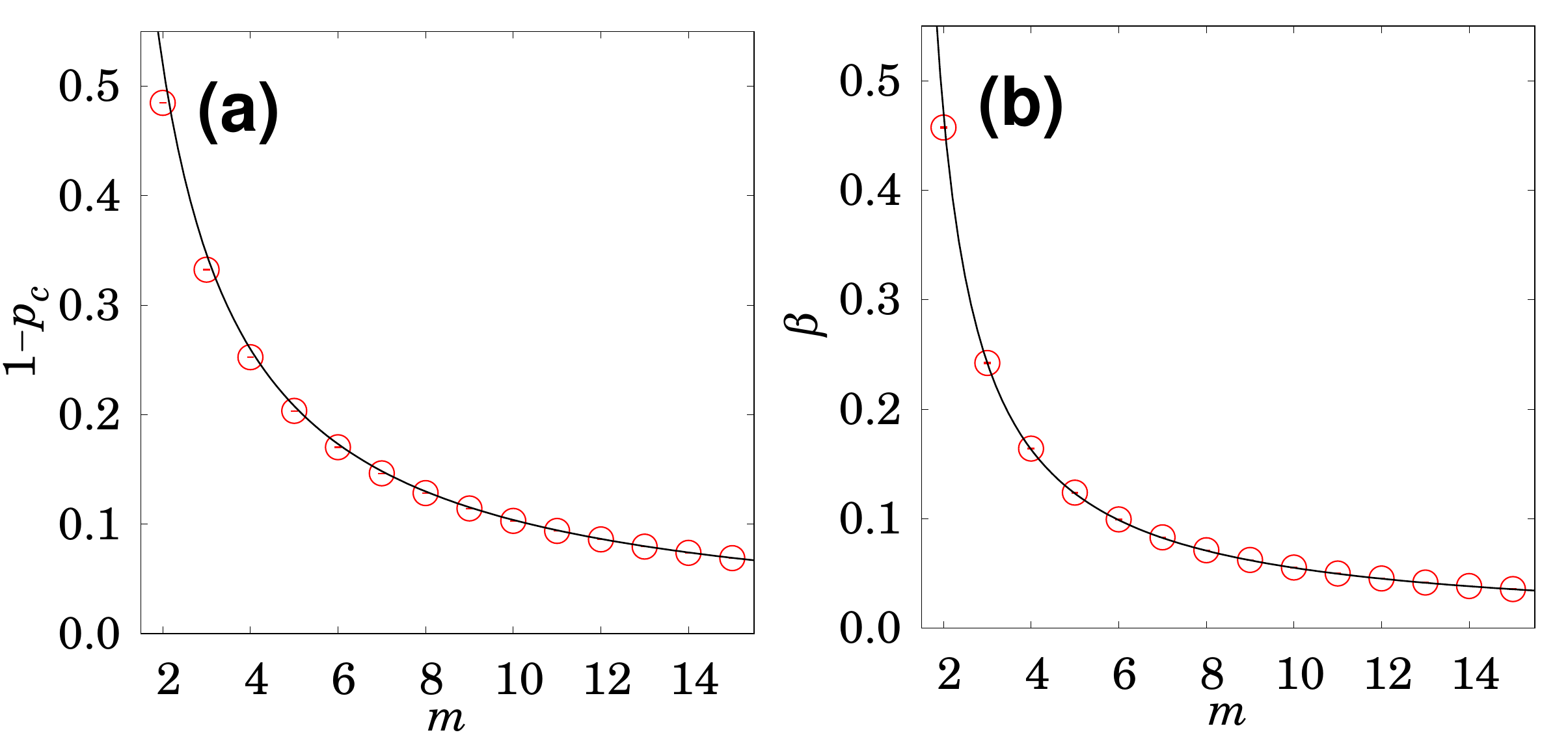}
\caption{(Color online) 
(a) Plot of $1-p_{c}$ versus $m$. Data points obtained from the rate equation seem to be fitted by the formula $1-p_{c}=1.04/m$. (b) Plot of $\beta$ versus $m$. Data points obtained from the rate equation are fitted by the formula $\beta=1/(2m-1.87)$. The critical exponent $\beta$ decreases algebraically with increasing $m$.
}
\label{fig:pc_and_beta_vs_m}
\end{figure}

Here, we check numerically the analytic result for the transition point $p_c$ and the scaling relations, and obtain various critical exponent values. To this end, we first obtain $P_m(s,p)$ from the rate equation Eq.~\eqref{eq:rate_eq_of_P_s_p} up to the order of $s^*$ explicitly~\cite{krapivsky_kinetic_2010}. Here, $s^*$ is taken as large as possible for numerical accuracy, but it should be less than $s_c$. Then, we determine the value of $p_c$ as that at which $P_m(s,p_c)$ exhibits power-law decay with respect to $s$~\cite{stauffer_introduction_2018,oh_explosive_2016, stauffer_scaling_1979}. This criterion is valid for a second-order percolation transition. Second, we determine the exponent $\tau$ by measuring the slope of $\ln P_m(s,p_c)$ with respect to $\ln s$, because the slope is $1-\tau$. Third, we determine the exponent $1/\sigma$ by plotting $P_m(s,p_c)s^{\tau-1}$ versus $s|p-p_c|^{1/\sigma}$ for different $p$ values. With an appropriate choice of $\sigma$, plots for different $p$ values can be collapsed onto a single curve. Next, to determine the exponent $\beta$, we plot $G(p)$ using $ 1-\sum_{s=1}^{s^{*}}P_m(s,p)$ versus $p-p_c$ ($p > p_c$) on the double logarithmic scale. We then measure the slope as $\beta$. Similarly, we obtain the values of the exponents $\gamma_P$ and $\gamma_Q$. 

\begin{figure*}[]
\includegraphics[width=1.0\linewidth]{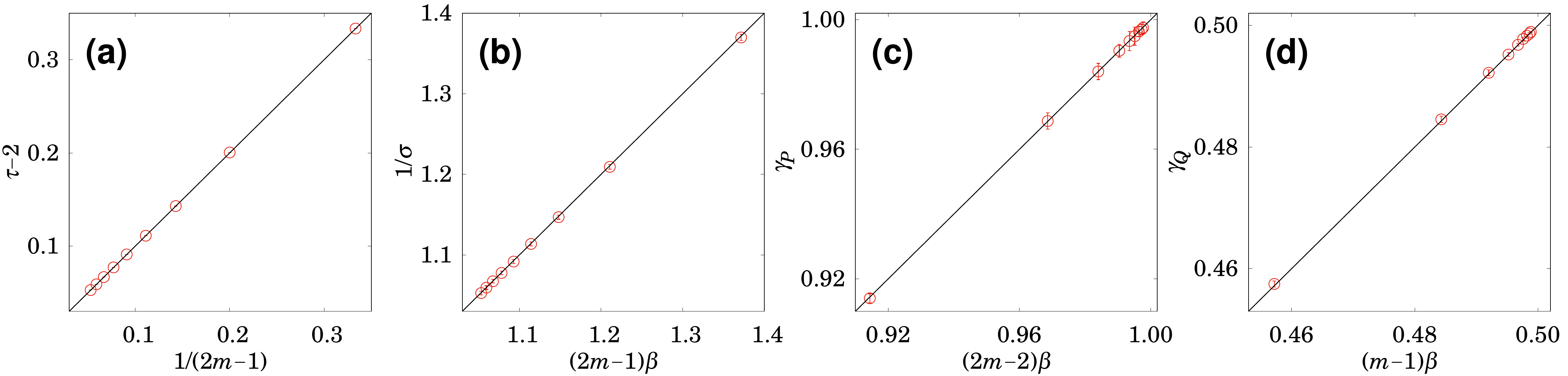}
\caption{(Color online) 
Test of the formulas for the exponents (a) $\tau=2+1/(2m-1)$, (b) $1/\sigma=(2m-1)\beta$, (c) $\gamma_P=2(m-1)\beta$, and (d) $\gamma_Q=(m-1)\beta$ using numerical data obtained from the rate equation. The error bars are presented in (a), (b), and (d). Data are fitted to the straight solid line, $y=x$, following Eqs.~\eqref{eq:scaling_rel_of_gamma_P}--\eqref{eq:scaling_rel_of_tau} in our growing network model.
}
\label{fig:growing_scaling_rel}
\end{figure*}

The estimated transition points and critical exponents for $m=2,\cdots,10$ are presented in Table~\ref{table_1}. We find that the transition point $p_{c}$ and exponents $\beta$ seem to behave as $1-p_{c}=1.04/m$ and $\beta=1/(2m-1.87)$, respectively, as shown in Fig.~\ref{fig:pc_and_beta_vs_m}. Moreover, the estimated values of the critical exponents $\tau$, $1/\sigma$, $\gamma_{P}$, and $\gamma_{Q}$ seem to satisfy the scaling relations in  Eqs.~\eqref{eq:scaling_rel_of_gamma_P}--\eqref{eq:scaling_rel_of_tau}, as shown in Fig.~\ref{fig:growing_scaling_rel}. 
\begin{table}[]
\centering
\caption{Values of the upper critical dimension $d_{u}$ obtained from the hyperscaling relation in growing networks for $m=2-5$. The correlation volume exponent $d_{u}\nu$ for the mean-field theory values $\nu = 1/2$ and numerically estimated $\bar{\nu}$ from the finite-size scaling approach is consistent with simulation data within errors. The network is grown to $N=2^{10} \times 10^{4}$, and the ensemble average is taken over more than $10^4$ samples for each $m$.}
\label{table_2}
\begin{tabular}{cccccc}
\hline\hline
~$m$~&~~$d_{u}$~~&~~~$d_{u}\nu$~~~&~~~$\bar{\nu}$~~~&~~~$\beta$~~~&~~~$1/\bar{\nu}$~~~ \\ \hline
2 &  5.66(30)  & 2.83(15)  & 2.61(7) & 0.458(38) & 0.383(10) \\
3 &  4.91(36)  & 2.46(18)  & 2.42(6)  & 0.243(30) & 0.413(10) \\
4 &  4.63(42)  & 2.32(21)  & 2.28(5)  & 0.165(27) & 0.439(10) \\
5 &  4.49(50)  & 2.24(25)  & 2.21(5)  & 0.124(25) & 0.452(10) \\ \hline
\end{tabular}
\end{table}
\begin{figure}[]
\center
\includegraphics[width=1.0\linewidth]{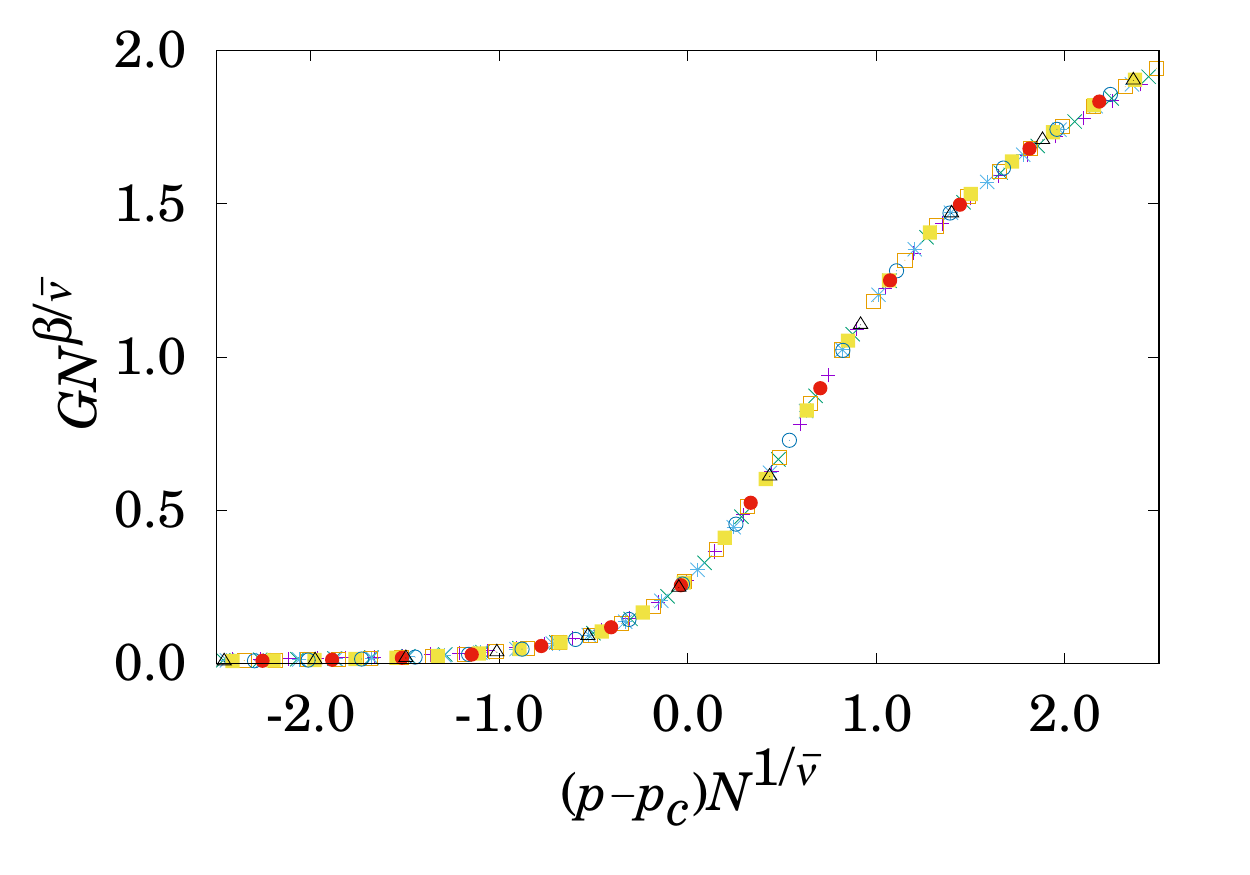}
\caption{(Color online) 
Finite-size scaling of $GN^{\beta/\bar{\nu}}$ versus $(p-p_c)N^{1/\bar{\nu}}$ from Monte Carlo simulations with more than $10^4$ realizations for system sizes $N/10^4 = 2^3, \cdots, 2^{10}$ for $m=2$. Data collapse onto a single curve with $p_c = 0.515(1)$, $1/\bar{\nu} = 0.383(10)$, and $\beta/\bar{\nu}=0.175(10)$, which give $\beta = 0.458(38)$.
}
\label{fig:Data_collase}
\end{figure}

We also check the critical exponents and transition point $p_c$ by direct simulations. We grow the networks to $N(t) = 2^{10} \times 10^4$ and repeat this growth more than $10^4$ times for $m=2$, as shown in Fig.~\ref{fig:Data_collase}. Using the finite-size scaling approach, $G_{N}(p)= N^{-\beta/\bar{\nu}}f\bigl((p-p_{c})N^{1/\bar{\nu}}\bigr)$~\cite{stauffer_introduction_2018,oh_explosive_2016,stauffer_scaling_1979}, we find $\beta/\bar{\nu}=0.175 \pm 0.010$, and $1/\bar{\nu} = 0.383 \pm 0.010$; thus $\beta = 0.458 \pm 0.038$ for $m=2$. Using $\nu = 1/2$ in the mean-field limit, the hyperscaling relation becomes $d_{u}-2=4\beta^{*}$ for a given $m$~\cite{da_costa_explosive_2010, da_costa_solution_2014}, where $d_{u}$ represents the upper critical dimension and $\beta^{*}$ is the critical exponent of the so-called observable order parameter. In our model, the observable order parameter is $G^{m}$ in the thermodynamic limit as $t \to \infty$,
because the probability that a node chosen under an aggregation rule belongs to a giant cluster acts as an observable order parameter. Our rule selects the node that belongs to the smallest of the $m$ candidate clusters; the probability that this node is in the giant cluster is $G^{m}$. Thus, the observable order parameter exponent is $\beta^{*}=m\beta$ in our model and the hyperscaling relation ultimately becomes $d_{u}-2=4m\beta$ for general values of $m$. For $m=2$, we obtain $d_u=5.66 \pm 0.30$ and the correlation volume exponent $d_{u}\nu=2.83 \pm 0.15$, which are consistent with the value obtained by simulations and the finite-size scaling approach, $d_u\nu=\bar{\nu}=2.61 \pm 0.07$, within errors. Therefore, the hyperscaling relation holds, and the upper critical dimensions in the growing network depend on $m$, but their numerical values differ from those of the static network model. Similarly, the hyperscaling relations are tested for different $m$ between 2 and 5, and the corresponding values of $d_{u}$ are presented in Table~\ref{table_2}. Finally, the analytical formula of $d_{u}$ is summarized in Table~\ref{table_3}.

\section{Summary and discussion}
\label{sec:summary and discussion} 
With regard to growing networks, we confirmed that the local suppression effect changed the type of percolation transition from infinite order to second order. Subsequently, we analytically derived the critical exponents $\tau$ and $1/\sigma$ for the probability of selecting a node in a cluster of size $s$, $P_m(s,p)\sim s^{1-\tau}f(s/s^*)$, where $s^*\sim (p-p_c)^{-1/\sigma}$ in terms of a control parameter $m$, representing the suppressing strength. Furthermore, transition point $p_c$ and other critical exponents were obtained in terms of $m$. They are summarized in Table~\ref{table_3} and compared with those in static networks~\cite{da_costa_explosive_2010, da_costa_solution_2014,oh_explosive_2016}. Our findings were confirmed by numerically solving the rate equations.

\begin{figure*}[]
\center
\includegraphics[width=1.0\linewidth]{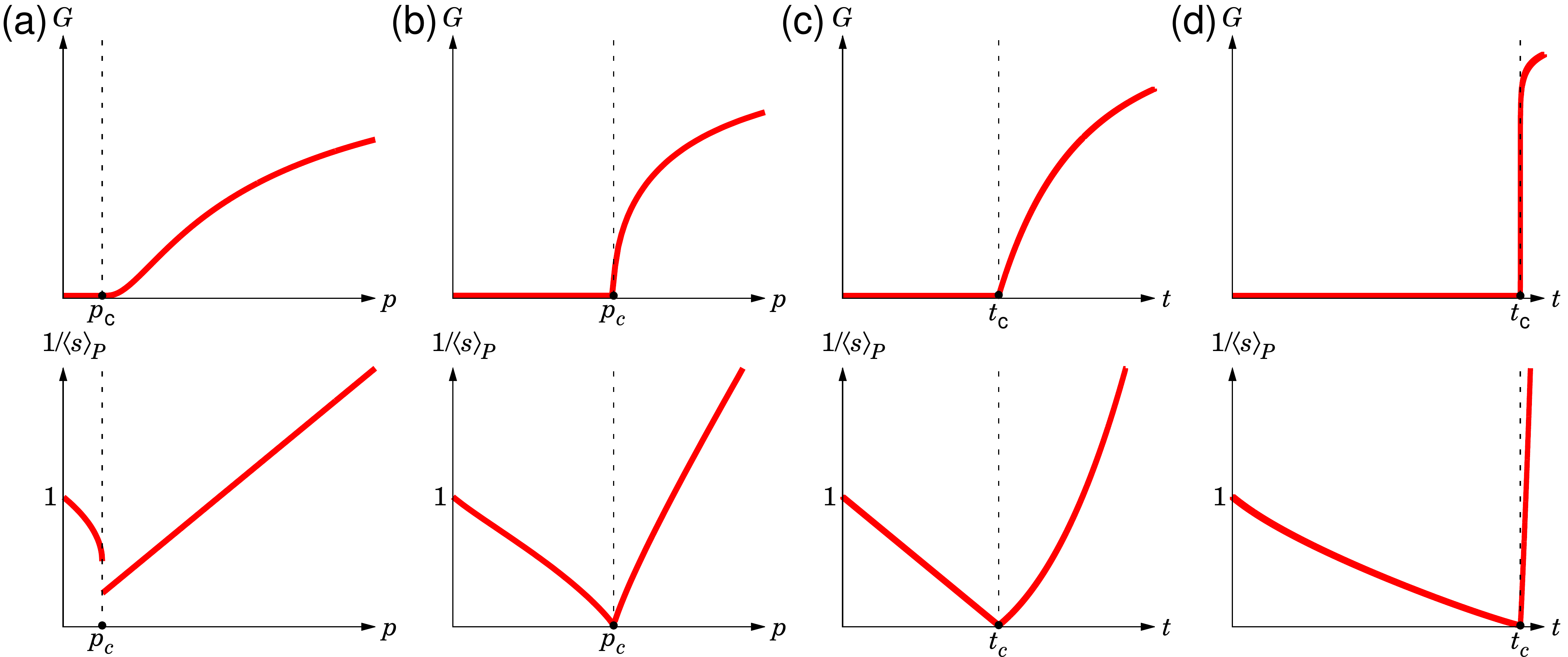}
\caption{(Color online) 
Schematic plots of the order parameter $G$ and the inverse first moment of $P_m(s,p)$, $1/\langle s \rangle_{P}$, for the (a) GRN, (b) $d$-GRN/$m$-GRN, (c) ER, and (d) $d$-ER/$m$-ER models. Schematic plots for the $m$-GRN and $m$-ER models are very similar to those for the $d$-GRN and $d$-ER models. The only difference between the two suppression rules is that twice as many nodes are selected under the da Costa rule as under the minimal rule. Thus, $m=2$ under the minimal rule and $m=1$ under the da Costa rule are reduced to the GRN model in growing networks and the ER model in static networks, respectively. 
}
\label{fig:Schematic_figure_comparison}
\end{figure*}

Interestingly, we discovered that as $m\to \infty$, the transition point and critical exponents behaved as $p_c\to 1$, $\tau\to 2$, $\sigma \to 1$, $\beta \to 0$, $\gamma \to 1$, and the upper critical dimension $d_u \to 4$. The fact that $\beta=0$ as $m \to \infty$ indicated that the percolation transition was discontinuous because the suppression effect became global~\cite{riordan_explosive_2011,cho_avoiding_2013}. We remark that, for growing networks, $p_c$ and the critical exponents algebraically approached their respective asymptotes as $m \to \infty$, whereas for static networks, they exponentially approached them as $m \to \infty$ and $d_u \to 2$. Accordingly, for a specified finite $m$, the order parameter $G$ increases slowly in growing networks, whereas it increases drastically in static networks, as shown in Fig.~\ref{fig:Schematic_figure_comparison}. 

\begin{table*}[]
\centering
\caption{Transition point $p_c$ ($t_c$); the order-parameter exponent $\beta$; the scaling relations of the critical exponents $\tau$, $1/\sigma$, and $\gamma_P$; and the upper critical dimension $d_u$ in growing (static) networks under two different suppression rules for candidate node selection. The minimal rule for the growing and static networks in Ref.~\cite{oh_explosive_2016} and the da Costa rule in the static networks in Refs.~\cite{da_costa_explosive_2010, da_costa_solution_2014} are compared. The difference between these two suppression rules is that twice as many nodes are selected under the da Costa rule as under the minimal rule.
}
\label{table_3}
\begin{tabular}{c|c|c|c|c|c|c|c}
\hline\hline

Network & Process & ~$p_{c}$ , $t_{c}$~ &~~~ $\beta$~~~ &~~~~~~ $\tau$~~~~~~ &~~~ $1/{\sigma}$~~~ &~~~~~ $\gamma_P$~~~~~  & ~~~$d_{u}$~~~  
\\ \hline

\multirow{2}{*}{\begin{tabular}[c]{@{}c@{}}Growing\\ network\end{tabular}} & \begin{tabular}[c]{@{}c@{}}minimal\\ rule~\cite{oh_explosive_2016} \end{tabular} & $1-\frac{1.81}{m}$ & $\frac{1}{m-1.56}$ & $2+\frac{1}{m-1}$ & $(m-1)\beta$ & $(m-2)\beta$  & $2+2m\beta$  
\\ \cline{2-8} 
                                 & \begin{tabular}[c]{@{}c@{}}da Costa \\ rule\end{tabular} & $1-\frac{1.04}{m}$ & $\frac{1}{2m-1.87}$ & $2+\frac{1}{2m-1}$ & $(2m-1)\beta$ & $2(m-1)\beta$ & $2+4m\beta$  
                                 \\ \hline
\multirow{2}{*}{\begin{tabular}[c]{@{}c@{}}Static\\ network\end{tabular}} & \begin{tabular}[c]{@{}c@{}}minimal\\ rule~\cite{oh_explosive_2016} \end{tabular} & $1-0.4e^{-0.59m}$ & $0.5e^{-0.70m}$ & $2+\frac{\beta}{1+(m-1)\beta}$ & $1+(m-1)\beta$ & $1+(m-2)\beta$ & $2+2m\beta$  
\\ \cline{2-8} 
                                & \begin{tabular}[c]{@{}c@{}}da Costa \\ rule~\cite{da_costa_explosive_2010, da_costa_solution_2014} \end{tabular} &  --  & $1.0e^{-1.43m}$ & $2+\frac{\beta}{1+(2m-1)\beta}$ & $1+(2m-1)\beta$ &  $1+2(m-1)\beta$ & $2 +4m\beta$  
                                \\ \hline
\end{tabular}
\end{table*}

The results we obtained in this study have led us to reinterpret the original results \cite{achlioptas_explosive_2009} regarding explosive percolation transitions from a new perspective. In the original paper, the AP was applied to static random networks under the product rule, where an edge minimizing the product of the sizes of merged components is selected between two selected random edges. The order parameter increased drastically even when only two candidate edges were used, which may correspond to $m = 2$ in our AP rules. Hence, the explosive percolation transition type was regarded as a discontinuous transition in the early stages. In retrospect, this hasty conclusion might have been made because the critical exponent $\beta$ of the order parameter decayed exponentially to zero with increasing $m$ for static networks, even though $\beta$ was still finite for a finite $m$. If the explosive percolation model had been considered with random growing networks, then such a conclusion would not have been made. 

\section*{Acknowledgement}
This research was supported by the National Research Foundation of Korea (NRF) through
Grant Nos. NRF-2014R 1A3A2069005 (B.K.) and NRF-2020R1A2C2010875 (S.W. S.), and by a TJ Park Science Fellowship from the POSCO TJ Park Foundation (S.W.S.).

\printcredits

\end{document}